\documentclass[numbered]{trbunofficial}
\usepackage{graphicx}
\usepackage{geometry}
\usepackage{array} 
\usepackage{float}
\usepackage{multirow}
\usepackage{amssymb}
\newcolumntype{P}[1]{>{\centering\arraybackslash}p{#1}}
\usepackage{tikz}
\newcommand*\circled[1]{\tikz[baseline=(char.base)]{
            \node[shape=circle,draw,inner sep=0.1pt, minimum size=0.2em] (char) {#1};}}
            
\usepackage{amsmath}
\usepackage{algorithm}
\usepackage{algpseudocode}
\usepackage[utf8]{inputenc}
\usepackage{array}
\usepackage{tabularx}
\usepackage{booktabs} 
\usepackage{multirow}
\usepackage{lscape} 
\usepackage{threeparttable}
\usepackage{verbatim} 
\usepackage{booktabs}
\usepackage{makecell}
\usepackage{colortbl}
\usepackage{xcolor}
\usepackage{soul} 
\sethlcolor{yellow}
\usepackage[hidelinks]{hyperref}

\AuthorHeaders{Yoon, Jeong, Lee, Kim, and Yoon}
\title{Integrating Urban Air Mobility with Highway Infrastructure\\: A Strategic Approach for Vertiport Location Selection in the Seoul Metropolitan Area}

\author{
  \textbf{Donghyun Yoon}\\
    Graduate School of Artificial Intelligence\\
    Korea Advanced Institute of Science and Technology, Daejeon, Republic of Korea, 34141 \\
    Email: tloveu49@kaist.ac.kr\\
  \hfill\break
    \textbf{Minwoo Jeong}\\
    Graduate School of Data Science\\
    Korea Advanced Institute of Science and Technology, Daejeon, Republic of Korea, 34141 \\
    Email: minwoo5003@kaist.ac.kr\\
  \hfill\break
  \textbf{Jinyong Lee}\\
    Graduate School of Innovation and Technology Management\\
    Korea Advanced Institute of Science and Technology, Daejeon, Republic of Korea, 34141 \\
    Email: liy35a@kaist.ac.kr\\
      \hfill\break
  \textbf{Seyun Kim}\\
    Department of Civil and Environmental Engineering\\
    Korea Advanced Institute of Science and Technology, Daejeon, Republic of Korea, 34141 \\
    Email: whataud@kaist.ac.kr\\
  \hfill\break
  \textbf{Yoonjin Yoon, Corresponding Author}\\
    Department of Civil and Environmental Engineering\\
    Korea Advanced Institute of Science and Technology, Daejeon, Republic of Korea, 34141 \\
    Email: yoonjin@kaist.ac.kr
}

\begin{document}
\maketitle

\section{Abstract}
This study focuses on identifying suitable locations for highway-transfer Vertiports to integrate Urban Air Mobility (UAM) with existing highway infrastructure. UAM offers an effective solution for enhancing transportation accessibility in the Seoul Metropolitan Area, where conventional transportation often struggle to connect suburban employment zones such as industrial parks. By integrating UAM with ground transportation at highway facilities, an efficient connectivity solution can be achieved for regions with limited transportation options. Our proposed methodology for determining the suitable Vertiport locations utilizes data such as geographic information, origin-destination volume, and travel time. Vertiport candidates are evaluated and selected based on criteria including location desirability, combined transportation accessibility and transportation demand. Applying this methodology to the Seoul metropolitan area, we identify 56 suitable Vertiport locations out of 148 candidates. The proposed methodology offers a strategic approach for the selection of highway-transfer Vertiport locations, enhancing UAM integration with existing transportation systems. Our study provides valuable insights for urban planners and policymakers, with recommendations for future research to include real-time environmental data and to explore the impact of Mobility-as-a-Service on UAM operations.

\hfill\break%
\noindent\textit{Keywords}: Urban Air Mobility, Vertiport, Intermodal Transfer Facility, Highway Facility, Seoul Metropolitan Area
\newpage

\section{1. Introduction} 
\begin{figure}[!ht]
  \centering
  \includegraphics[width=0.95\textwidth]{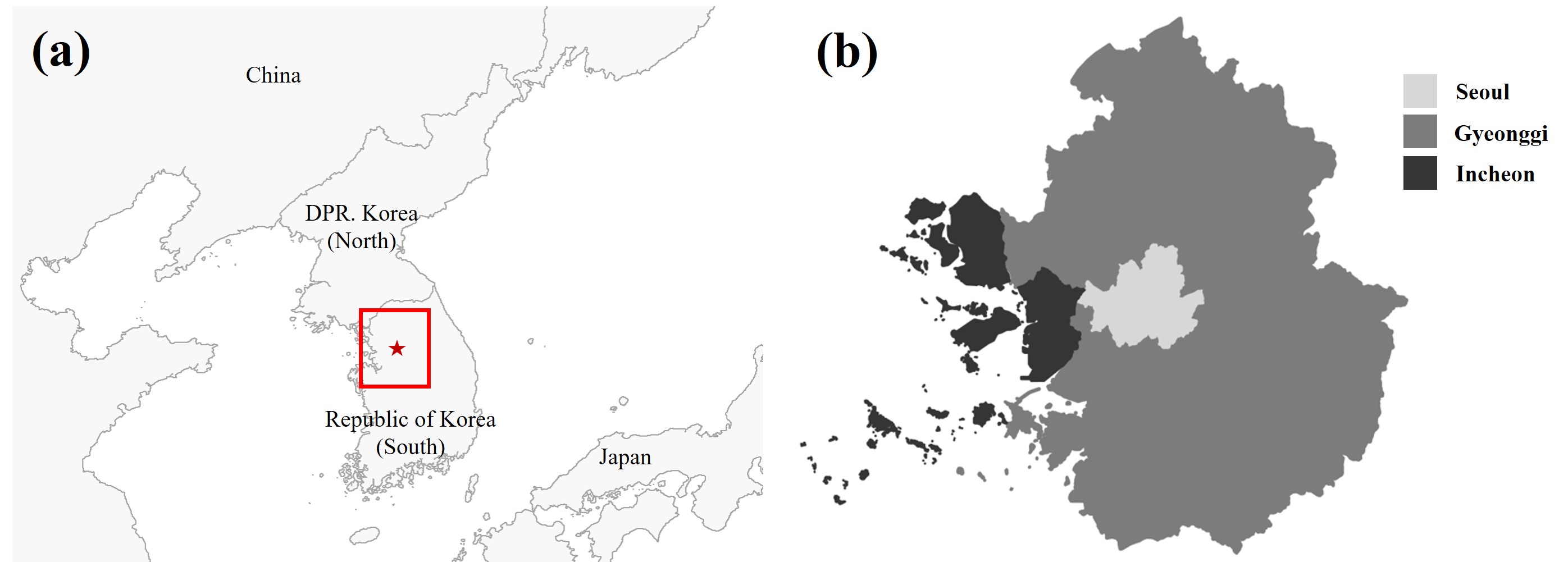}
  \caption{(a) The Location of the Seoul Metropolitan Area in South Korea (b) The Location of Seoul, Gyeonggi and Incheon in the Seoul Metropolitan Area}\label{fig1}
\end{figure}

The City of Seoul is the South Korea's hub for the economic, culture, politics and technology. Spanning 234 square miles, the city's population density is around 41,000 per square mile, which is more than 1.5 times the density of New York City (around 27,480 per square mile) and higher than that of Tokyo (around 16,480 per square mile) \cite{KoreanPopulationDensity, USPopulationDensity, TokyoPopulationDensity}. Since early 2000s, the city gradually lost population to neighboring regions such as Gyeonggi Province and Incheon City, which are part of the Seoul Metropolitan Area 
 (Figure \ref{fig1}), as shown in Table \ref{tab:Statistics_metro}.
However, expansion of Seoul into the metropolitan area was driven not only by the demand for more affordable housing but also by the relocation of various businesses to Gyeonggi Province and Incheon in the form of so-called Industrial Parks. As of 2024, Gyeonggi Province and Incheon City host more than 3,525,818 companies including major global companies such as Samsung, Hyundai and LG \cite{KoreanNumCompanies}.

\begin{table}[H]
    \begin{center}
        \caption{Population and Industrial Park Statistics in the Seoul Metropolitan Area \cite{population, industrial_park}}\label{tab:Statistics_metro}
    \begin{tabular}{>{\raggedright\arraybackslash}m{0.1\linewidth}>{\raggedright\arraybackslash}m{0.15\linewidth}>{\raggedright\arraybackslash}m{0.15\linewidth}>{\raggedright\arraybackslash}m{0.1\linewidth}>{\raggedright\arraybackslash}m{0.1\linewidth}>{\raggedright\arraybackslash}m{0.1\linewidth}>{\raggedright\arraybackslash}m{0.1\linewidth}}
    \toprule[1.5pt]

\textbf{Region} & \multicolumn{2}{c}{\textbf{Population}} & \multicolumn{2}{c}{\textbf{Total Area of}} & \multicolumn{2}{c}{\textbf{Number of}} \\
& & & \multicolumn{2}{c}{\textbf{Industrial Parks ($mi^2$)}} & \multicolumn{2}{c}{\textbf{Industrial Parks}} \\
\cmidrule(lr){2-3} \cmidrule(lr){4-5} \cmidrule(lr){6-7}
& \centering\textbf{2001} & \centering\textbf{2023} & \centering\textbf{2001} & \centering\textbf{2023} & \centering\textbf{2001} & \textbf{2023} \\
\midrule[1.5pt]
Seoul & 10,091,137 & 9,400,365 & 1.50 & 1.27 & 2 & 4 \\
\hline
Gyeonggi & 9,451,079 & 13,781,261 & 19.18 & 96.77 & 49 & 192 \\
\hline
Incheon & 2,550,919 & 3,008,826 & 4.73 & 8.46 & 6 & 16 \\
    \bottomrule[1.5pt]
\end{tabular}
\label{table:data}
\end{center}
\end{table}

The demand for travel within the Seoul Metropolitan Area has naturally increased. However, despite the various and frequent transportation options available for traveling to Seoul, accessibility to these industrial complexes from other regions remains notably restricted. For instance, according to Google Maps, traveling from Busan to Seoul (about 330km) by express train takes about 2 hours and 50 minutes . In contrast, the shorter distance from Busan to the A.W. National Industrial Park (about 295km) involves three modes of transportation and takes 4 hours and 20 minutes. To address these such challenges in accessibility, Urban Air Mobility (UAM) can provide access to areas that are difficult to reach with conventional ground transportation. Moreover, efficient transfers at major transportation hubs such as highways can enhance connectivity.
Successful implementation requires integrating UAM with ground transportation, enhancing overall efficiency and complementing current infrastructure \cite{jiang2024}.
In South Korea, Express-HUB(EX-HUB) is an essential part of the highway infrastructure to serve bus transfer passengers, especially for inter-city buses \cite{han2016seoul}. 
Integrating UAM with ground transportation at highway facilities such as EX-HUB and rest areas can offer an efficient solution for accessing areas with limited transportation options.

The rest of paper is organized as follows. Chapter 2 presents the literature review. Chapter 3 introduces methodology, followed by analysis results in Chapter 4 and analysis discussions in Chapter 5. Finally, Chapter 6 provides conclusions and future research directions. 

\section{2. Literature Review} 
\subsection{2.1 Vertiport Location Selection}
Vertiports, which are essential infrastructure for UAM, should be installed in appropriate locations considering demand, economic feasibility, operational capability, and safety. Numerous studies employing various methodologies have been conducted to address these considerations.

The multi-criteria decision-making (MCDM) method involves making optimal choices by assessing multiple evaluation criteria across different options, employing techniques such as Analytic Hierarchy Process (AHP) and Focus Group Interviews (FGI). Fadhil (2018) analyzed the factors of UAM ground infrastructure placement and suitability using WLC and AHP techniques \cite{fadhil2018gis}. Lee et al. (2023) proposed an optimal Vertiport and UAM network by calculating various topographic, population and social data of Seoul and its weights through the FGI technique for pilots \cite{lee2023}. Recent studies have adequately considered whether the area is permissible for flight. 

From a data-driven perspective, recent studies on Vertiport location assessment have utilized data from four major categories—spatial, mobility, environmental, and socio-economic—for conducting multifaceted analyses. Straubinger et al. (2021) evaluated the UAM ecosystem in different urban spatial structures using population density data \cite{straubinger2021_employment}. Bulusu et al. (2021) analyzed feasible combinations of Vertiport locations using traffic data \cite{bulusu2021}. Kotwicz Herniczek et al. (2022) evaluated the impact of airspace restrictions on the feasibility of Vertiports and UAM routes using airspace data \cite{kotwicz_airspace}. EASA (2021) analyzed the social acceptance of UAM in Europe using surveys and noise assessments \cite{easa2021study}. Other related literature is summarized in Table \ref{tab:Major Data}.

\subsection{2.2 Integration of UAM with Existing Transportation Systems}

Since the emergence of the UAM concept, researchers have proposed various approaches for integrating Vertiports into the existing transportation system. Fadhil (2018) utilized GIS to select locations for UAM among ground infrastructure locations \cite{fadhil2018gis}. Rajendran et al. (2019) proposed Vertiport locations that enhance airport accessibility by analyzing taxi data in New York City \cite{rajendran2019}. Wang et al. (2023) suggested utilizing UAM to transfer passengers from suburban areas to airports, thereby facilitating access to conventional air transportation through an on-demand transfer service \cite{wang2023}. Those studies have focused on transfers between UAM and existing urban transportation networks, and the idea of linking highway infrastructure with UAM has not been proposed.

\begin{table}[H]
    \caption{List of Data Used for Vertiport Location Assessment}\label{tab:Major Data}
        \begin{center}
        \begin{tabular}{>{\raggedright\arraybackslash}m{0.2\linewidth}>{\raggedright\arraybackslash}m{0.22\linewidth}>{\raggedright\arraybackslash}m{0.5\linewidth}} \toprule[1.5pt]
            \multicolumn{1}{c}{\textbf{Category}} & \multicolumn{1}{c}{\textbf{Data}} & \multicolumn{1}{c}{\textbf{Application}} \\ 
            \midrule[1.5pt]
            \multirow{4}{*}[-0.5ex]{Spatial Data} & Population Density \cite{straubinger2021_employment, macias2023integrated} & Potential demand evaluation of Vertiport location based on population distribution and density Information \\ \cline{2-3}
             & Employment Density \cite{straubinger2021_employment, daskilewicz2018progress} & Evaluation of access to key CBDs based on employment density information \\ \cline{2-3}
             & Land Use \cite{yedavalli_land, vascik2019development} & Selection of locations based on the status of land use, such as commercial districts, residential districts, and industrial districts \\ \cline{2-3}
             & Height of Building \cite{lee2023, vascik2020geometric} & Evaluation of safety during take-off and landing through building data in the city center \\ \hline
            \multirow{4}{*}[-0.5ex]{Mobility Data} & Traffic \cite{bulusu2021, rajendran2023capacitated} & Evaluation of accessibility by road traffic, bus and subway passengers \\ \cline{2-3}
             & Commuting Pattern \cite{rimjha_commuter, murcca2021identification} & Identification of key user groups by analyzing commuting flow over time \\ \cline{2-3}
             & Public Transport \cite{wu2021_network, lim2019selection}& Evaluation of accessibility by analyzing bus, subway line and stop locations \\ \cline{2-3}
             & Traffic Congestion \cite{bulusu2021, jin2024robust} & Evaluation of traffic congestion level and to derive the optimal location \\ \hline
            \multirow{4}{*}[-0.5ex]{Environmental Data} & Air Space \cite{kotwicz_airspace, vascik2020geometric} & Selection of information-based locations related to aviation regulations, such as flight prohibitions/restrictions/control zones \\ \cline{2-3}
             & Nature Reserve \cite{lee2023, chen2022scalable} & Evaluation of environmental constraints such as Nature Reserve and Waterfront Areas \\ \cline{2-3}
             & Noise \cite{rimjha_noise, ison2023analysis} & Evaluation of the noise level from flying UAM vehicles and Identification of areas where the noise levels is higher than criteria. \\ \cline{2-3}
             & Weather \cite{bernyk2023aerodynamic, park2022comparison} & Analysis of historical wind data for UAM takeoff and landing direction assessment \\ \cline{2-3}
             & Cost of Construction \cite{taylor2020design, rimjha2021urban} & Economic feasibility assessment based on cost information for the Vertiport construction \\ \hline
            \multirow{4}{*}[-0.5ex]{\raggedright \makecell{Socio-Economic \\ Data}} & Cost of Operation \cite{tarafdar_cost, mendonca2022advanced} & Long-term operational feasibility assessment based on cost information required to operate Vertiport \\ \cline{2-3}
             & Estimation Earnings \cite{kai2022, murcca2021identification} & Evaluation of economic feasibility by predicting revenue from Vertiport operations \\ \cline{2-3}
             & Income Level \cite{cho2022uam, daskilewicz2018progress} & Economic feasibility assessment of UAM services through income distribution information \\ \cline{2-3}
             & Social Acceptance \cite{easa2021study, ison2024consumer}& Evaluation of social impact and acceptability based on survey or noise impact assessment \\ \cline{2-3}
             & Related Regulations \cite{perperidou_regul, vascik2020geometric} & Evaluation of regulations, policies, and restrictions for selecting Vertiport locations
             \\ 
             \bottomrule[1.5pt]
        \end{tabular}
    \end{center}
\end{table}

\section{3. Methodology}
\begin{figure}[!ht]
  \centering
  \includegraphics[width=1\textwidth]{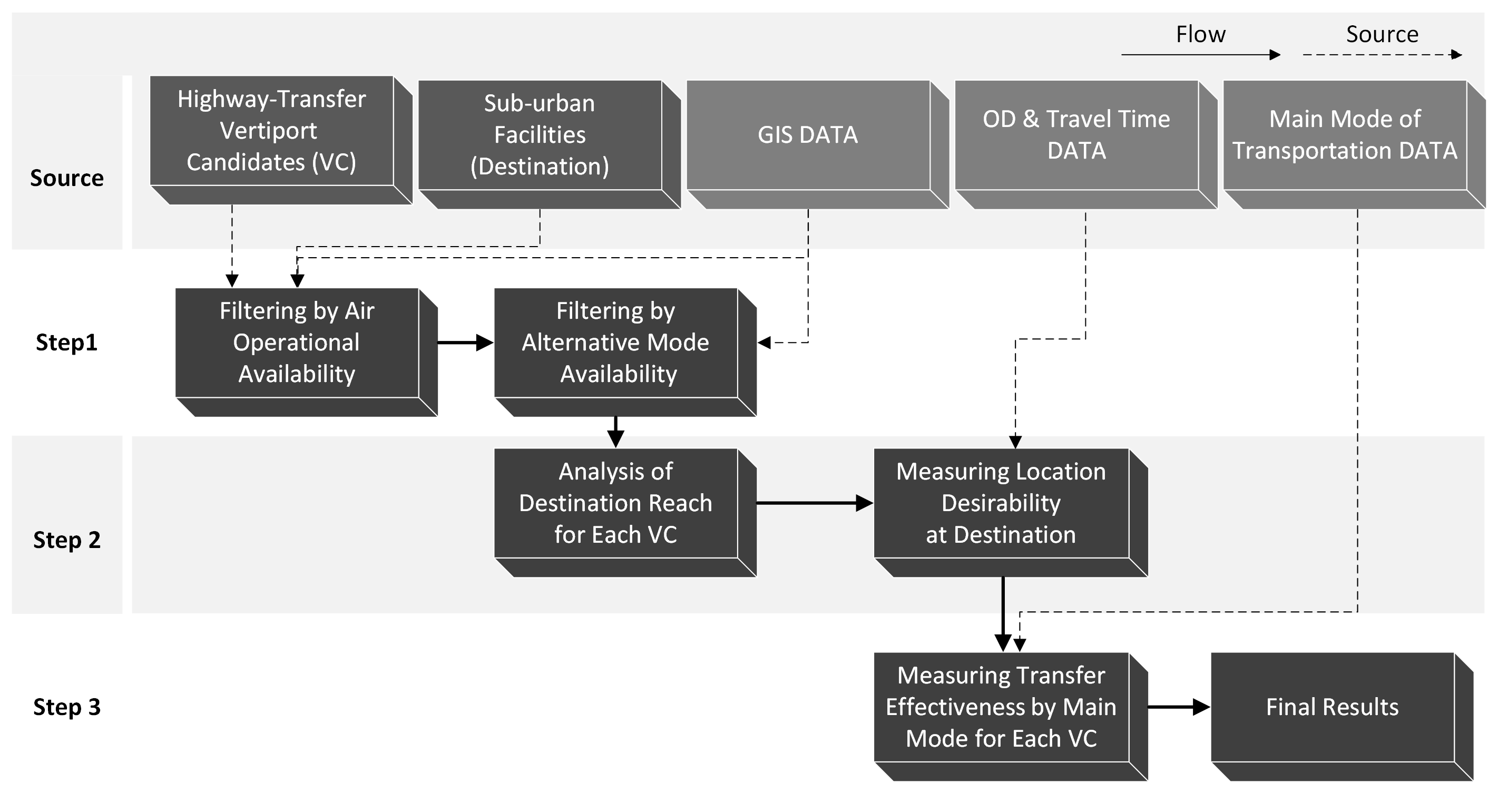}
  \caption{Methodology for Determining Highway-transfer Vertiport Candidates}\label{figure:2}
\end{figure}

This study aims to propose a methodology for determining the suitable locations for highway-transfer Vertiports on highway facilities. It is crucial to assess both the feasibility of air operations and the connectivity to existing transportation networks when identifying appropriate Vertiport sites. Furthermore, evaluating the vulnerability and demand of the transportation network is essential to enhance public welfare \cite{wang2024vulnerability}. Consequently, we present a methodology for selecting the suitable locations by considering these key factors, as illustrated in Figure \ref{figure:2}.

The methodology involves assessing both the vulnerability and demand for public transportation within UAM service areas. One's travel time to the destination will vary depending on which combination of routes  and modes are selected. This variation in travel time serves as a measure of the traveler's efficiency. To apply this methodology, factors such as transportation mode, accessibility, demand, and location desirability are considered. 

\subsubsection{1) Transportation Modes}
The Transportation modes utilized in this study can be categorized into three types. The bus is the main mode of transportation to reach a highway-transfer Vertiport. UAM is the mode of transportation to travel from the highway-transfer Vertiport to the Vertiport in destination. The alternative mode of transportation refers to the means of travel used when UAM is unavailable, substituting travel from the highway-transfer Vertiport such as taxi or subway.

\subsubsection{2) Transportation Accessibility}
The transportation accessibility refers to the possibility to reach a specific destination from the starting point using available transportation. Primarily, travel time, distance, and travel cost are measured and used as indicators (Hansen (1959) \cite{hansen1959accessibility}). In this study, it is assumed that the longer the travel time, the lower the accessibility, using Travel Time as the primary indicator of destination accessibility.

\subsubsection{3) Transportation Demand}
The transportation demand refers to the willingness to pay for the benefits of transportation services at the individual or social level. Generally, an increase in transportation demand is associated with a corresponding increase in traffic volume. Origin-Destination (OD) volume data are used as an indicator of demand for the destination. In this study, it is assumed that the greater the OD volume accessing the destination, the higher the demand for transportation.

\subsubsection{4) Location Desirability}
The location desirability is an evaluation index that indicates the suitability of a site as a UAM operation destination. Higher location desirability implies greater suitability for UAM operations, corresponding to an operation destination by utilizing transportation accessibility and demand. Longer travel times, indicating lower transportation accessibility, correspond to higher location desirability. Similarly, greater OD volumes, indicating higher transportation demand, also correspond to higher location desirability. The reason for considering only two variables to measure location desirability is that they directly impact the public welfare aspect of transportation when selecting a location. Transportation accessibility evaluates the vulnerability of public transportation networks, while transportation demand assesses the potential for UAM demand. Therefore, the study aims to simplify the analysis, enhance the existing transportation network, and clearly address public welfare concerns by reflecting boarding demand.
\\

Table \ref{tab:KeyNotation} provides a detailed explanation of all notations used throughout the methodology.

\begin{table}[H]
    \caption{Nomenclature}\label{tab:KeyNotation}
    \begin{center}
    \renewcommand{\arraystretch}{1}
    \begin{tabular}{>{\centering\arraybackslash}m{0.09\linewidth} >{\raggedright\arraybackslash}m{0.85\linewidth}} \toprule[1.5pt]
        \multicolumn{1}{c}{\textbf{Not.}} & \multicolumn{1}{c}{\textbf{Description}} \\ 
        \midrule[1.5pt]
        \( \mathbf{F} \) & Facility grid, where $f_{ij}$ is 1 if a region contains a highway facility, considered a Vertiport candidate, otherwise 0. \\ \hline
        \( \mathbf{C} \) & Constraint grid, where $c_{ij}$ represents the cumulative presence of constraints; $c_{ij} > 0$ indicates at least on constraint, and $c_{ij}$ is 0 denotes a constraint-free cell \\ \hline
        \( \mathbf{S} \) & Selected grid, initialized as \( \mathbf{S} = \mathbf{F} \) and updated iteratively; cells corresponding to non-zero values in \( \mathbf{F} \) are set to 0 during filtering \\ \hline
        \( \mathbf{x}^{Time}_{i,j} \) & Travel time of the \( j \)-th public transportation node to the \( i \)-th destination. \\ \hline
        \( \mathbf{x}^{OD}_{i,j} \) & Origin-Destination volume of the \( j \)-th public transportation node to the \( i \)-th destination. \\ \hline
        \( \tilde{\mathbf{x}}^{Time}, \tilde{\mathbf{x}}^{OD} \) & Min-max scaled values of Travel Time and OD volume. \\ \hline
        \( \gamma \) & Weight in the convex combination. \\ \hline
        \( \mathbf{x}^{Score}_i \) & Score for the \( i \)-th destination. \\ \hline
        \( \mathbf{x}^{\text{numbus}}_v \) & Number of bus routes passing through the \( v \)-th Vertiport candidate. \\ \hline
        \( l_v \) & Number of destinations included in the \( v \)-th Vertiport candidate. \\ \hline
        Score\(_v\) & Score for the \( v \)-th Vertiport candidate. \\
        \bottomrule[1.5pt]
    \end{tabular}
    \end{center}
\end{table}

\subsection{3.1 Filtering by Air Operation Availability}
The initial step involves analyzing the availability of air operations for both the Vertiport candidate sites on highway infrastructure and the designated destinations. To evaluate the suitability of air operations at each candidate site, a grid system is established, and major constraints to UAM operations are identified based on relevant regional flight regulations. Grids containing such constraints are then filtered out. This systematic approach ensures that only feasible sites, free from significant operational hindrances, are considered for further evaluation.

\subsubsection{3.1.1 Grid Setup and Filtering Process}
\begin{figure}[H]
  \centering
  \includegraphics[width=1\textwidth]{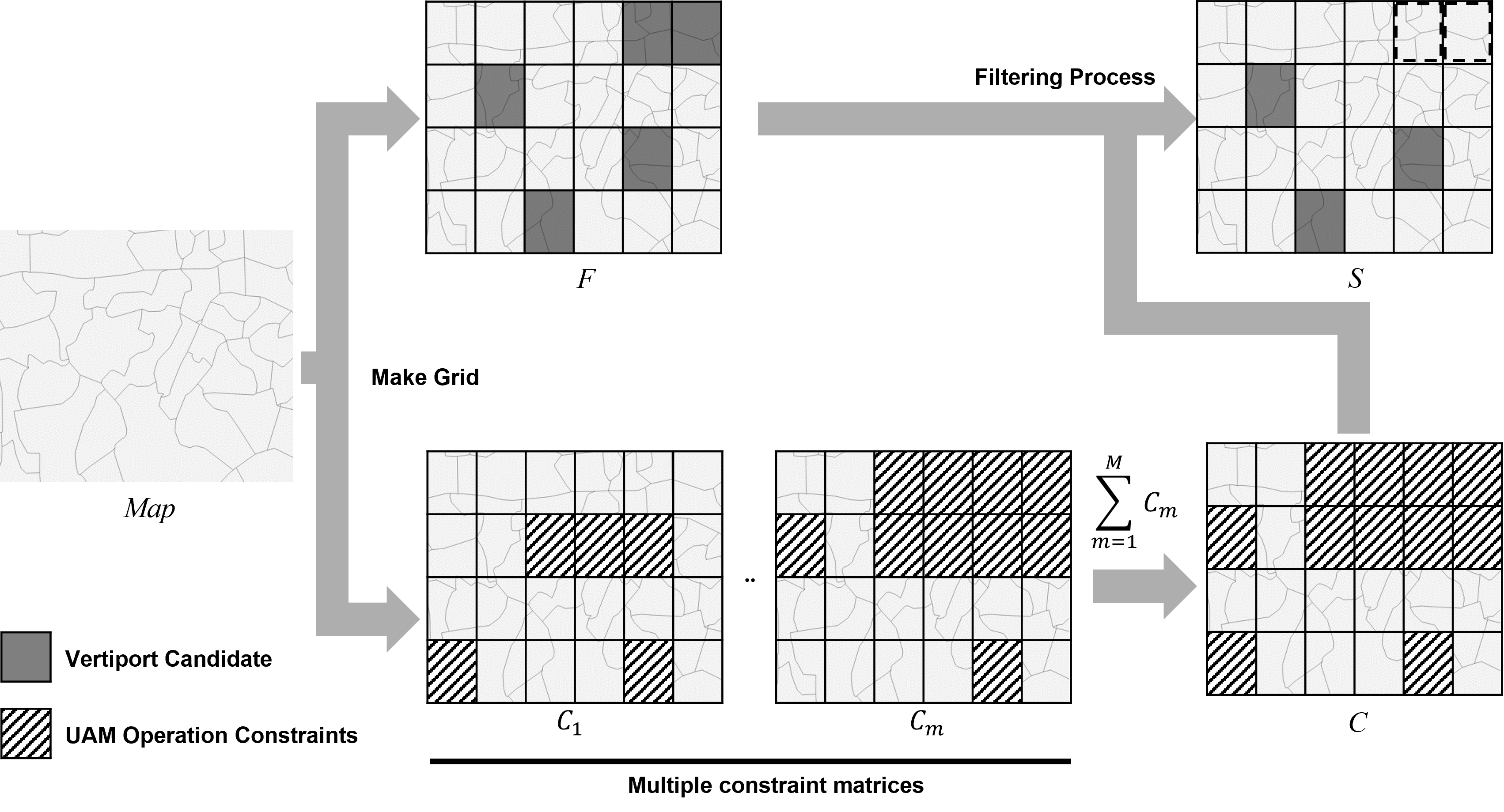}
  \caption{Illustration of Grid Setup and Filtering Process}\label{figure:3}
\end{figure}

Assuming that the evaluation area is discretized as a $P \times Q$ grid, we define three grid matrices $\mathbf{F}$, $\mathbf{C}$, and $\mathbf{S}$ as follows:
\\
\\
\begin{linenomath}
\(\mathbf{F}(i,j) \in \mathbb{R}^{P \times Q}\) is the facility grid matrix where
\begin{equation}
\mathbf{F}(i,j) =
\begin{cases} 
1, & \text{if grid } (i,j) \text{ contains a vertiport candidate site,} \\
0, & \text{otherwise.}
\end{cases}
\end{equation}
\end{linenomath}
\\
\\
\begin{linenomath}
Constraint Grid \(\mathbf{C}(i,j) \in \mathbb{R}^{P \times Q}\) is an integer matrix where the value of grid \((i,j)\) denotes the total presence of operational constraints. To elaborate, \(\mathbf{C}(i,j)\) is the matrix summed over multiple constraint matrices \(\mathbf{C}_m(i,j)\), where 
\begin{equation}
\mathbf{C}_m(i,j) =
\begin{cases} 
1, & \text{if grid } (i,j) \text{ contains a constraint of category } m, \\
0, & \text{otherwise.}
\end{cases}
\end{equation}
\end{linenomath}
\\
\\
\begin{linenomath}
In this study, we considered 8 constraint categories including Prohibited Area, Restricted Area, Danger Zone, Military Operational Area, Control Zone, Aerodrome Traffic Zone, Alert Area, and Terrain Obstacles, yielding \(\mathbf{C}(i,j) = \sum_{m=1}^8 \mathbf{C}_m(i,j)\).
\end{linenomath}
\\
\\
\begin{linenomath}
Finally, $\mathbf{S}(i,j) \in \mathbb{R}^{P \times Q}$ is the Selected Grid where
\begin{equation}
\mathbf{S}(i,j) =
\begin{cases} 
1, & \text{if } \mathbf{F}(i,j) = 1 \text{ and } \mathbf{C}(i,j) = 0, \\
0, & \text{otherwise.}
\end{cases}
\end{equation}
\end{linenomath}
\\
\\
In other words, $\mathbf{S}(i,j)$ is a binary matrix where only the grids with candidate site with no operational constraint is 1. In practice, $\mathbf{S}$ is initialized as $\mathbf{S} = \mathbf{F}$ and iteratively updated through a filtering process(algorithm \ref{algorithm1} and Figure \ref{figure:3}) to incorporate each constraint category step by step.

\begin{algorithm}
\caption{Filtering Process}
\label{algorithm1}
\begin{algorithmic}[1]
\For {each cell $(x, y)$ in \( \mathbf{F} \)}
    \If {$f_{xy} = 1$ and $c_{xy} = 0$}
        \State $s_{xy} \gets 1$
    \Else 
        \State $s_{xy} \gets 0$
    \EndIf
\EndFor
\end{algorithmic}
\end{algorithm}

\subsection{3.2 Filtering by Alternative Mode of Transportation Availability}
\begin{figure}[H]
  \centering
  \includegraphics[width=1\textwidth]{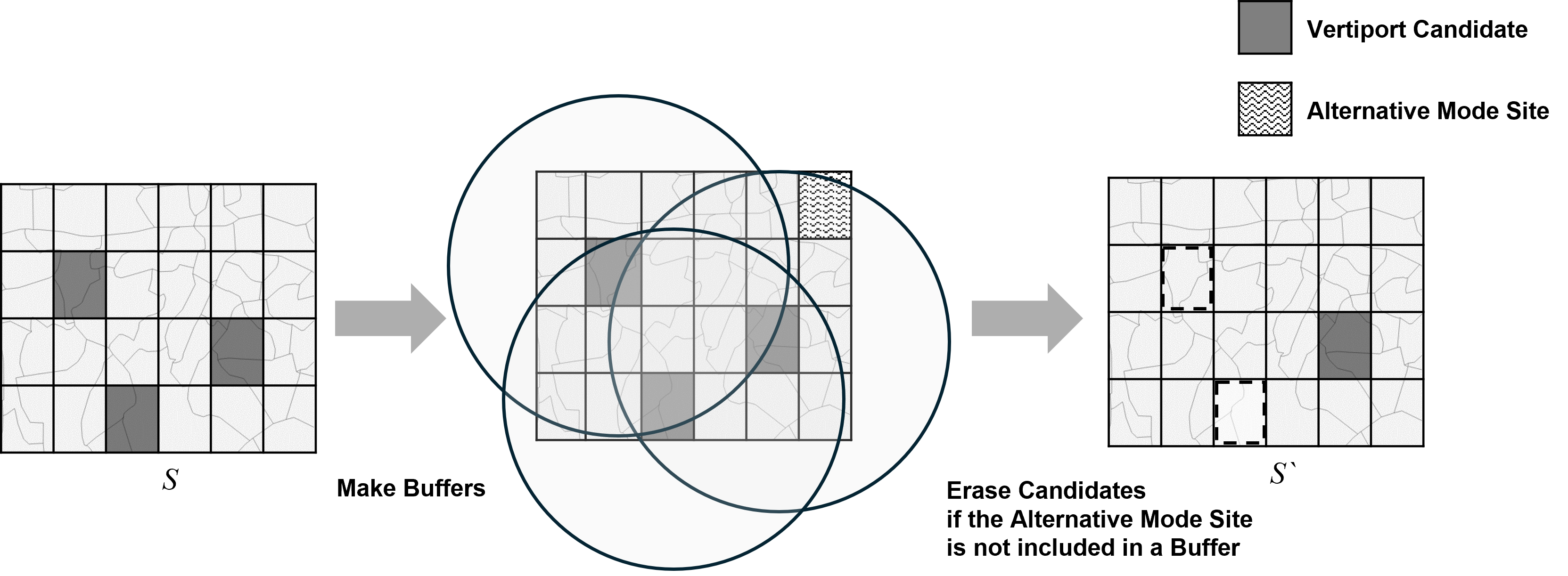}
  \caption{Illustration of Filtering by Alternative Mode of Transportation Availability}\label{figure:4}
\end{figure}

The availability of alternative mode of transportation options is analyzed for Vertiport candidates. If UAM operations are suddenly disrupted due to local weather or visibility issues, it is crucial to have alternative mode of transportation available. A filter is applied to the GIS to identify whether there are alternative mode of transportation sites within a certain buffer zone around each candidate. The size of the buffer zone is determined according to the regulations of the area. After applying the filter, the final Vertiport candidates (\( \mathbf{S'} \)) are selected (Figure \ref{figure:4}).

\subsection{3.3 Measuring Location Desirability by Score Function}
The location desirability of the Vertiport candidates is measured based on the existing transportation network vulnerability at the destination. In this process, \circled{1} a destination that can be reached within a certain distance from the Vertiport candidate site is measured, and \circled{2} a score function evaluating the accessibility and demand of the destination's public transportation network is applied to measure the location desirability.

\subsubsection{3.3.1 Analysis of Destination Reach for Each Vertiport Candidate}
The analysis determines whether multiple destinations can be reached from each Vertiport candidate. In this methodology, constraints are considered instead of merely measuring a straight travel path between the origin and destination, and an optimal travel path is determined through a route-finding algorithm. The operational coverage of each Vertiport candidate can then be identified.

\subsubsection{3.3.2 Score Measurement}
A score function is utilized to evaluate the accessibility and demand of the destination's public transportation network, which in turn is used to measure location desirability. Assume there are \( N \) destinations and \( M \) public transportation nodes, such as bus stops, subway stations, and train stations, where \( M \) refers to the nearest public transportation nodes to the destinations. Each node is represented by indices \( i \) and \( j \). \( i \) is defined as the index for destinations, where \( i = 1, 2, \ldots, N \). \( j \) is defined as the index for public transportation nodes, where \( j = 1, 2, \ldots, M \). \( v \) is defined as the index for Vertiport candidates, where \( v = 1, 2, \ldots, V \). \( \mathbf{x}_{i,j} \) represents the \( \mathbf{x} \) value of the \( j \)-th public transportation node in the \( i \)-th destination.
The variable \( \mathbf{x}_{i,j}^{Time} \) is defined as the travel time of the \( j \)-th public transportation node in the \( i \)-th destination. The total values for travel time are defined as follows:

\begin{linenomath}
    \begin{equation}
        \begin{aligned}
            \mathbf{x}_{i, \cdot}^{\text{Time}} &= \sum_{j=1}^{M} \mathbf{x}_{i,j}^{\text{Time}}
        \end{aligned}
    \end{equation}
\end{linenomath}
\\
The variable \( \mathbf{x}_{i,j}^{OD} \) is defined as Origin-Destination volume of the \( j \)-th public transportation node in the \( i \)-th destinations. The total values for OD volume are defined as follows:

\begin{linenomath}
    \begin{equation}
        \begin{aligned}
            \mathbf{x}_{i, \cdot}^{\text{OD}} &= \sum_{j=1}^{M} \mathbf{x}_{i,j}^{\text{OD}}
        \end{aligned}
    \end{equation}
\end{linenomath}
\\
Equations \eqref{eq:xt} and \eqref{eq:xod} show the min-max scaled total values (denoted with a tilde) for Travel Time and OD volume corresponding to the \(i\)-th destination. The max and min of \( \mathbf{x}^{T, OD} \) mean the maximum and minimum values among those included in the group of \( i \) destinations.

\begin{linenomath}
    \begin{equation}
        \begin{aligned}
        \tilde{\mathbf{x}}_{i, \cdot}^{\text{Time}} &= \frac{\mathbf{x}_{i, \cdot}^{\text{Time}} - \mathbf{x}_{\min}^{\text{Time}}}{\mathbf{x}_{\max}^{\text{Time}} - \mathbf{x}_{\min}^{\text{Time}}} \quad & 
        \end{aligned}
        \label{eq:xt}
    \end{equation}
\end{linenomath}
\\
\begin{linenomath}
    \begin{equation}
        \begin{aligned}
        \tilde{\mathbf{x}}_{i, \cdot}^{\text{OD}} &= \frac{\mathbf{x}_{i, \cdot}^{\text{OD}} - \mathbf{x}_{\min}^{\text{OD}}}{\mathbf{x}_{\max}^{\text{OD}} - \mathbf{x}_{\min}^{\text{OD}}}
        \end{aligned}
        \label{eq:xod}
    \end{equation}
\end{linenomath}
\\
The score for the \( i \)-th destination is defined as a convex combination of the scaled total values. The weight of Origin-Destination demand and Travel Time is set equally when \( \gamma \) is 0.5, but \( \gamma \) can be set differently depending on the intention of the policymaker using this methodology.

\begin{linenomath}
    \begin{equation}
        \begin{aligned}
        \mathbf{x}_{i}^\text{Score} = \gamma \tilde{\mathbf{x}}_{i, \cdot}^{\text{Time}} + (1 - \gamma) \tilde{\mathbf{x}}_{i, \cdot}^{\text{OD}} \quad \text{where } 0 \leq \gamma \leq 1 \text{ (default: } \gamma = 0.5\text{)}
        \end{aligned}
        \label{eq:score_function}
    \end{equation}
\end{linenomath}

\subsection{3.4 Measuring Transfer Effectiveness by Main Mode of Transportation}
\begin{figure}[H]
  \centering
  \includegraphics[width=1\textwidth]{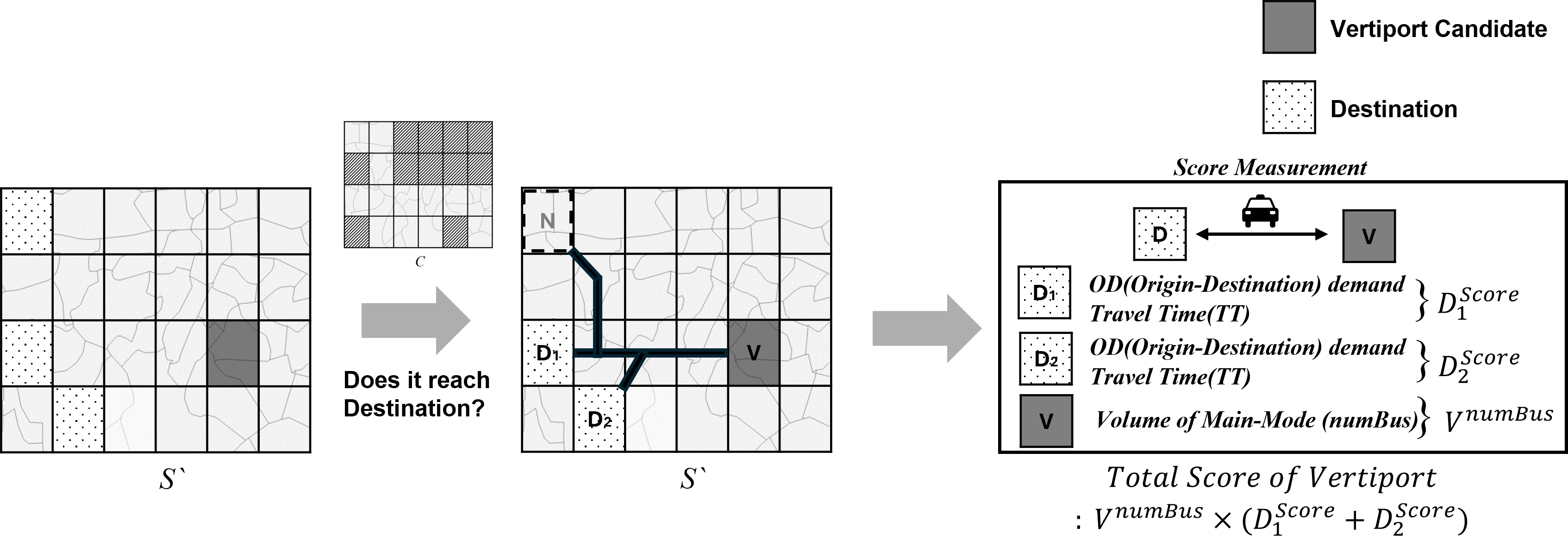}
  \caption{Illustration of Measuring Location Desirability and Transfer Effectiveness}\label{figure:5}
\end{figure}

The effectiveness of the main mode of transportation at each Vertiport candidate is measured based on the assumption that more frequent or greater volumes of transportation passing through a Vertiport candidate lead to an increase in the number of UAM transfers. Let \( v \) be the index for Vertiport candidates, where \( v = 1, \ldots, V \). Each Vertiport candidate contains a set of destinations indexed by \( i \). Let \( I_v \) be the number of destinations included in the \( v \)-th Vertiport candidate. The score for the \( v \)-th Vertiport candidate is measured by multiplying the volume of main mode of transportation by the sum of the scores of the destinations (Figure \ref{figure:5}). In this study, the volume of main mode of transportation was assumed to be the number of bus routes (\(numBus\)) passing through the Vertiport candidate.

\begin{linenomath}
    \begin{equation}
        \begin{aligned}
        \text{Score}_v = \mathbf{x}_{v}^\text{numBus} \times \sum_{i=1}^{I_v} \mathbf{x}_{i}^\text{Score}
        \end{aligned}
    \end{equation}
\end{linenomath}

\newpage
\section{4. Analysis}
\subsection{4.1 Analysis Setup}
We conducted an analysis focusing on the Seoul Metropolitan Area. In this study, 148 Vertiport candidates were selected based on the facilities (Toll Gate, Rest Area, EX-HUB) installed on the highway, and buses were selected for main mode of transportation. Among these, EX-HUB is a compound word of Expressway and Hub, referring to an integrated transfer facility designed to connect highways with public transportation in Korea. Additionally, We selected national industrial parks as destination. For business travelers, UAM has the advantages of \circled{1} saving travel time by avoiding traffic congestion and \circled{2} enabling fast movement through 3D public spaces even in areas with weak land transportation networks. In this context, the assumption that national industrial parks in suburban areas will generate sufficient demand for business travelers is plausible. Accordingly, data were selected, collected and processed as shown in Table ~\ref{tab:tab2}.

\begin{table}[H]
    \caption{List of Data Used in the Analysis}\label{tab:tab2}
    \begin{center}
        \begin{tabular}{>{\raggedright\arraybackslash}m{0.15\linewidth}>{\raggedright\arraybackslash}m{0.35\linewidth}>{\raggedright\arraybackslash}m{0.12\linewidth}>{\raggedright\arraybackslash}m{0.25\linewidth}} 
        \toprule[1.5pt]
            \multicolumn{1}{c}{\textbf{Step}} & \multicolumn{1}{c}{\textbf{Data Name}} & \multicolumn{1}{c}{\textbf{Format}} & \multicolumn{1}{c}{\textbf{Source}} \\ 
            \midrule[1.5pt]
            \multirow{6}{=}{Basic Environment} 
             & City/County Boundary & SHP & Goverment of South Korea (MOLIT) \\ \cline{2-4}
             & Standard Node Link & SHP & ITS Center \\ \cline{2-4}
             & Nationwide Rest Area Data & CSV & Korea Expressway Corp \\ \cline{2-4}
             & EX-HUB (Current/Planned) & CSV & Self Crawling \\ \cline{2-4}
             & Bus Terminal Status & SHP & Gyeonggi Province \\ \hline
            \multirow{7}{=}{Step 1} 
             & Prohibited Zone & \multirow{7}{=}{SHP} & \multirow{7}{=}{Goverment of South Korea (MOLIT) } \\ \cline{2-2}
             & Restricted Zone &  & \\ \cline{2-2}
             & Danger Zone & &  \\ \cline{2-2}
             & Control Area &  &  \\ \cline{2-2}
             & Aerodrome Traffic Zone &  &  \\ \cline{2-2}
             & Alert Area &  &  \\ \cline{2-2}
             & Digital Elevation Model (DEM) &  &  \\ \hline
            \multirow{2}{=}{Step 2} 
             & TMAP API & API & SKT \\ \cline{2-4}
             & Taxi OD Traffic Volume & CSV & Korea Transport Inst \\ \hline
            \multirow{1}{=}{Step 3} 
             & Traffic Network GIS DB & SHP & Korea Transport Inst \\ 
             \bottomrule[1.5pt]
        \end{tabular}
    \end{center}
\end{table}

\subsection{4.2 Filtering by Air Operation and Alternative Mode of Transportation Availability}
\begin{figure}[H]
  \centering
   \includegraphics[width=1\textwidth]{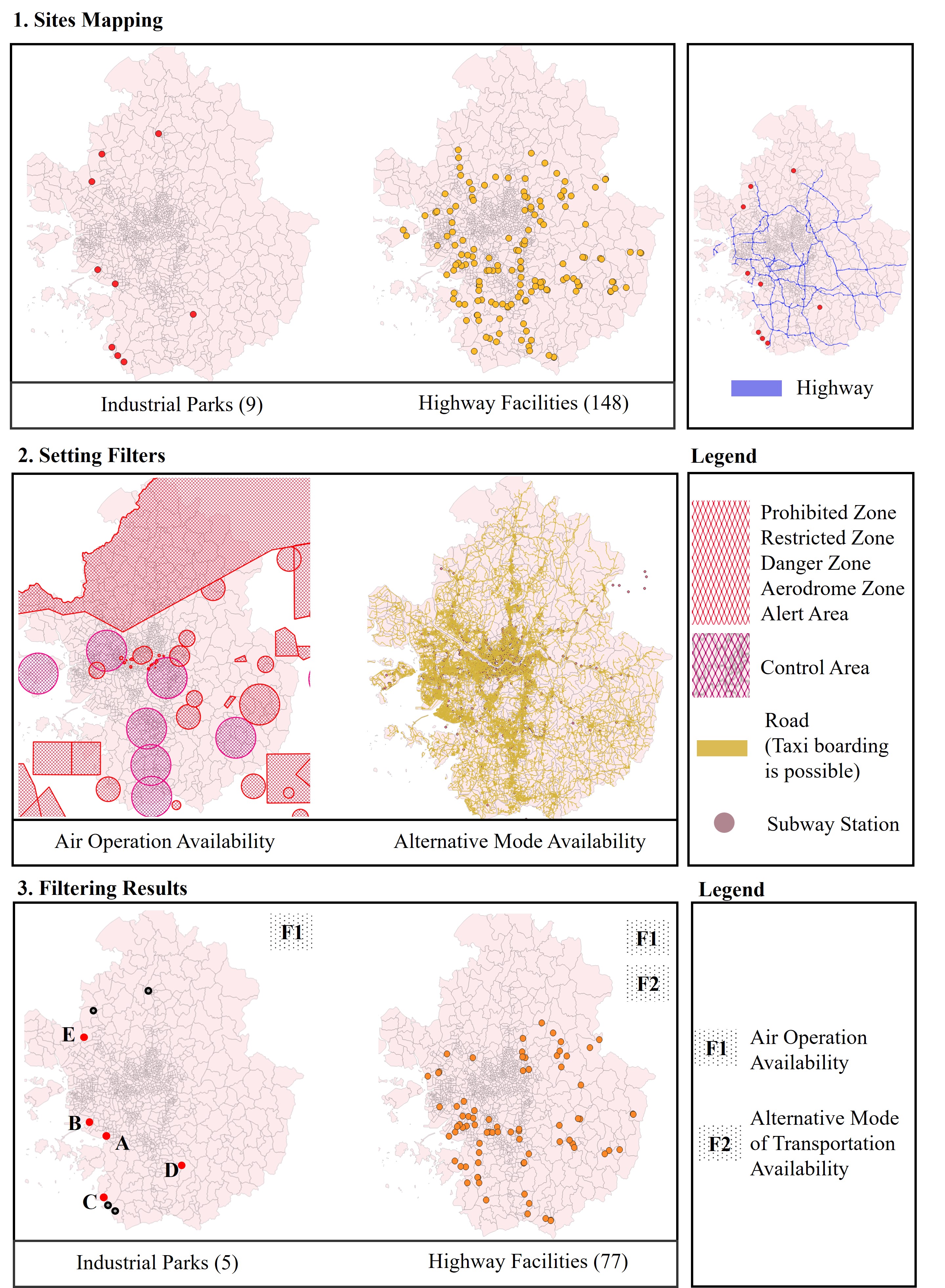}
  \caption{Result of Filtering Process(Air Operation Availability, Alternative Mode of Transportation Availability)}\label{figure:6}
\end{figure}

Figure \ref{figure:6} shows the assessment of Vertiport candidate's availability for air operation and alternative mode of transportation. UAM operational constraints, national industrial parks and highway facilities were mapped on the Seoul Metropolitan Area, and air operation availability and alternative mode of transportation availability filters were applied to identify operable areas. The air operation availability was assessed based on Table \ref{tab:major components}. The components of this table referenced the air space management regulation\cite{Korea_Airspace_Management_Regulations} and UAM ConOps \cite{U_UAMCONOPS, K_UAMCONOPS}. The alternative mode of transportation availability was assessed by checking the proximity to taxi-accessible roads or subway stations within a 450-meter radius \cite{Korea_Transfer_Center}. As a result, 5 candidates were selected from 9 national industrial parks, and 77 candidates were selected from 148 highway facilities.

\begin{table}[H]
    \caption{List of UAM Operational Constraints} \label{tab:major components}
    \begin{center}
        \begin{tabular}{>{\raggedright\arraybackslash}m{0.25\linewidth}>{\raggedright\arraybackslash}m{0.35\linewidth}>{\raggedright\arraybackslash}m{0.3\linewidth}} 
        \toprule[1.5pt]  
            \multicolumn{1}{c}{\textbf{Constraints}}& \multicolumn{1}{c}{\textbf{Definition}}& \multicolumn{1}{c}{\textbf{UAM Flight Restriction}}\\ 
            \midrule[1.5pt] 
            Prohibited area& Designated to prohibit the flight of aircraft to protect critical national facilities or to prevent unauthorized border crossings by aircraft. & \multirow{2}{=}{UAM operation within the area is completely prohibited for national security and public safety}\\ \cline{1-2}
            Restricted area& Designated to restrict flights in order to protect aircraft from dangers such as anti-aircraft attacks or for other specific reasons.& \\ \hline  
            Danger Zone& Designated to alert the flight where there is an anticipated risk to aircraft or ground facilities when aircraft are in flight.& Restrictions due to collisions with ground facilities and risks of various accidents (due to low visibility, building wind, etc.)\\ \hline  
            Military Operational Area& Designated to alert the flight where non-dangerous military flight activities or military operations are carried out, such as air combat maneuvering, air interception, and low-altitude tactics of military aircraft& Occurrence of irregular restrictions such as elevation and route separation, strict flight permits required, NOTAM\\ \hline  
            Control Zone& Designated to manage the operation of flight around the airport where air traffic control is installed& Operation through a specific entry/exit point when attempting to pass through the area\\ \hline  
            Aerodrome Traffic Zone& Designated to provide traffic information in non-control grade D airspace & UAMs in the area are only provided with limited traffic information for watch flights\\ \hline  
            Alert Area& Designated to alert the flight where large-scale pilot training or abnormal forms of aviation activities are carried out& Flight permit required, training and abnormal aviation activities result in aircraft collisions and path changes\\ \hline  
            Terrain obstacles& Topographical obstacles that are impossible to operate due to invasion beyond the altitude at which the UAM operates & Elevation and visibility restrictions on UAM routes, emergency landing restrictions, and deterioration of weather conditions (e.g. mountain turbulence)\\ 
            \bottomrule[1.5pt] 
        \end{tabular}
    \end{center}
\end{table}

\subsection{4.3 Calculating Location Desirability by Score Function}
We applied the equation \eqref{eq:score_function} to calculate the location desirability of the previously selected Vertiport candidates. First, we created buffer zones to measure whether UAM can reach national industrial parks from highway facilities, assuming a one-way trip distance of 30km. We used air operation availability filters to create a 100m x 100m grid that accounts for constraints. Subsequently, the JPS (Jump Point Search) algorithm \cite{harabor2012jps} was employed to identify accessible candidate sites considering constraints in the process. Next, we calculated the location desirability score for each national industrial park. 

In this study, taxi data were used to calculate accessibility and demand. In areas with weak public transportation networks, the available public transportation options are insufficient or inconvenient, making taxis an important alternative mode of transportation. Considering these characteristics, the taxi travel time calculated by the navigation API was used to measure accessibility, and taxi OD volume data was used to measure demand. The taxi OD volume data were collected from weekdays in March, June, September, and December of 2022, with the analysis timeframes set to Morning Peak, Evening Peak, and Off-Peak Hours. Public transportation nodes were assumed to be Bus, Rail, and Subway. We calculated the OD value and travel time from the nearest node to each park to determine their location desirability. The final location desirability score for each park was calculated using Equation \eqref{eq:score_function}, as shown in Table \ref{tab:national_industrial_park} below.

\begin{table}[H]
    \caption{Location Desirability Score for National Industrial Park}\label{tab:national_industrial_park}
    \begin{center}
        \begin{tabular}{>{\raggedright\arraybackslash}m{0.3\linewidth}>{\raggedright\arraybackslash}m{0.1\linewidth}>{\raggedright\arraybackslash}m{0.1\linewidth}>{\raggedright\arraybackslash}m{0.1\linewidth}>{\raggedright\arraybackslash}m{0.1\linewidth}>{\raggedright\arraybackslash}m{0.12\linewidth}}
            \toprule[1.5pt]
            \multicolumn{1}{c}{\textbf{Industrial Park $(i)$}} & \multicolumn{1}{c}{\textbf{$\mathbf{x}_{i, \cdot}^{\text{Time}}$}} & \multicolumn{1}{c}{\textbf{$\tilde{\mathbf{x}}_{i, \cdot}^{\text{Time}}$}} & \multicolumn{1}{c}{\textbf{$\mathbf{x}_{i, \cdot}^{\text{OD}}$}} & \multicolumn{1}{c}{\textbf{$\tilde{\mathbf{x}}_{i, \cdot}^{\text{OD}}$}} & \multicolumn{1}{c}{\textbf{$\mathbf{x}_{i}^{\text{Score}}$}} \\ 
            \midrule[1.5pt]
            A(Banwol) & 71.696 & 0.15314 & 1107 & 0.283921 & 0.21853 \\ \hline
            B(Sihwa) & 60.89 & 0 & 3788 & 1 & 0.5 \\ \hline
            C(Asan Wojeong) & 131.452 & 1 & 80 & 0.009615 & 0.504808 \\ \hline
            D(Yongin) & 72.081 & 0.158603 & 44 & 0 & 0.079301 \\ \hline
            E(Paju Publishing) & 73.346 & 0.176523 & 199 & 0.0414 & 0.108961 \\ 
            \bottomrule[1.5pt]
        \end{tabular}
    \end{center}
\end{table}

\subsection{4.4 Calculating Transfer Effectiveness}
The effectiveness of the main mode of transportation at each highway facility was analyzed. Each highway facility encompasses a set of accessible industrial parks. The location desirability of each accessible industrial park was aggregated and multiplied by the volume of main mode of transportation for each highway facility to calculate its final score.

\subsection{4.5 Results}
The total result are shown in Table \ref{tab:Results2}. In Table \ref{tab:Results2}, out of the 56 total candidates, 43 are Toll Gates, 11 are Rest Areas, and 2 are EX-HUBs. Compared to the initial analysis settings, it was found that 42\% of Toll Gates, 28\% of Rest Areas, and 25\% of EX-HUBs were selected. Among the 56 total candidates, the average number of bus routes is 63.3, with 74\% of the total candidates having fewer bus routes than this average. 

\begin{table}[H]
    \centering
    \caption{Results of Final Candidates for Highway-Transfer Vertiport}
    \begin{center}
    \label{tab:Results2}
    \begin{tabular}{>{\raggedright\arraybackslash}m{0.14\linewidth}>{\raggedright\arraybackslash}m{0.08\linewidth}>{\raggedright\arraybackslash}m{0.09\linewidth}>{\raggedright\arraybackslash}m{0.075\linewidth}>{\raggedright\arraybackslash}m{0.14\linewidth}>{\raggedright\arraybackslash}m{0.08\linewidth}>{\raggedright\arraybackslash}m{0.09\linewidth}>{\raggedright\arraybackslash}m{0.075\linewidth}} 
    \toprule[1.5pt]
        \multicolumn{1}{c}{\textbf{$v$}} & \multicolumn{1}{c}{\textbf{$\mathbf{x}_{v}^\text{numBus}$}} & \multicolumn{1}{c}{\textbf{$\sum \mathbf{x}_{i}^\text{Score}$}} & \multicolumn{1}{c}{\textbf{${Score}_v$}} & \multicolumn{1}{c}{\textbf{$v$}} & \multicolumn{1}{c}{\textbf{$\mathbf{x}_{v}^\text{numBus}$}} & \multicolumn{1}{c}{\textbf{$\sum \mathbf{x}_{i}^\text{Score}$}} & \multicolumn{1}{c}{\textbf{${Score}_v$}} \\
        \midrule[1.5pt]
        E.Gunpo T\tnote{1} & 213 & 0.71853 & 153.04 & Anseong T & 65 & 0.079301 & 5.15 \\ \hline
        Guseong E\tnote{2} & 478 & 0.297831 & 142.36 & Gonjiam T & 63 & 0.079301 & 4.99 \\ \hline
        Dongcheon E & 477 & 0.297831 & 142.06 & W.Suji T & 16 & 0.297831 & 4.76 \\ \hline
        Gunja T & 190 & 0.71853 & 136.52 & S.Anseong T & 5 & 0.71853 & 3.59 \\ \hline
        Ansan R\tnote{3} & 125 & 0.71853 & 89.81 & Balan T & 3 & 0.723338 & 2.17 \\ \hline
        Siheung T & 101 & 0.71853 & 72.57 & W.Anseong T & 24 & 0.079301 & 1.90 \\ \hline
        Jukjeon (S) R & 241 & 0.297831 & 71.77 & Cheongbuk T & 2 & 0.723338 & 1.44 \\ \hline
        Mado T & 54 & 1.223338 & 66.06 & Anseong (M.P) R & 16 & 0.079301 & 1.26 \\ \hline
        Songsan Mado T & 54 & 1.223338 & 66.06 & Bibong T & 1 & 1.223338 & 1.22 \\ \hline
        Joam T & 54 & 1.223338 & 66.06 & Anseong (M.J) R & 11 & 0.079301 & 0.87 \\ \hline
        Hwaseong (M) R & 53 & 1.223338 & 64.83 & S.Anseong T & 3 & 0.237903 & 0.23 \\ \hline
        Hwaseong (S) R & 53 & 1.223338 & 64.83 & W.Icheon T & 3 & 0.237903 & 0.23 \\ \hline
        N.Suwon T & 81 & 0.797831 & 64.62 & Goyang T & 0 & 0.108961 & 0.00 \\ \hline
        W.Seoul T & 75 & 0.71853 & 53.88 & S.Gwang myeong T & 0 & 0.71853 & 0.00 \\ \hline
        W.Ansan T & 67 & 0.71853 & 48.14 & S.Gunpo T & 0 & 0.71853 & 0.00 \\ \hline
        Siheung Sky R & 61 & 0.71853 & 43.83 & S.Bibong T & 0 & 1.223338 & 0.00 \\ \hline
        W.Siheung T & 49 & 0.71853 & 35.20 & Docheok T & 0 & 0.079301 & 0.00 \\ \hline
        Geumjeong E & 34 & 0.71853 & 24.43 & Dongtan T & 0 & 0.079301 & 0.00 \\ \hline
        Anseong (S) R & 292 & 0.079301 & 23.15 & Munhak Tunnel T & 0 & 0.71853 & 0.00 \\ \hline
        Maesong T & 17 & 1.223338 & 20.79 & Mulwang T & 0 & 0.71853 & 0.00 \\ \hline
        Uiwang T & 23 & 0.797831 & 18.35 & W.Yongin T & 0 & 0.079301 & 0.00 \\ \hline
        Icheon (N) R & 174 & 0.079301 & 13.79 & Shihwa T & 0 & 0.71853 & 0.00 \\ \hline
        Icheon (H) R & 174 & 0.079301 & 13.79 & Yeonseong T & 0 & 0.71853 & 0.00 \\ \hline
        Bugok T & 17 & 0.797831 & 13.56 & Ilsan Br. T & 0 & 0.108961 & 0.00 \\ \hline
        S.Incheon T & 17 & 0.71853 & 12.21 & Jungri T & 0 & 0.079301 & 0.00 \\ \hline
        New Airport T & 17 & 0.608961 & 10.35 & Cheongna T & 0 & 0.608961 & 0.00 \\ \hline
        Yeongjong Br. R & 17 & 0.608961 & 10.35 & Hwaseong T & 0 & 1.223338 & 0.00 \\
        \bottomrule[1.5pt]
    \end{tabular}
    \begin{tablenotes}
        \item[ ] \textit{T : Toll Gate, E : EX-HUB, R : Rest Area, ${Score}_v$ values are truncated to three decimal places}
    \end{tablenotes}
    \end{center}
\end{table} 
\section{5. Discussion}
\subsection{5.1 Characteristics of Top Candidates(in Each Type of Highway Facility)}
The top candidates for each type of Highway Facility are listed in Table \ref{tab:highway_scores}. E. Gunpo T was identified as the top Toll Gate, Ansan R as the top Rest Area, and Guseong E as the top EX-HUB. While E. Gunpo T and Ansan R serve the same industrial parks, their results differed due to the number of buses, with Ansan R having approximately 60\% fewer buses. Despite both being on the Yeongdong Highway, Ansan R is situated further west of the Metropolitan area centroid, resulting in fewer bus routes. Guseong E is notable for having the highest number of buses but a relatively low location desirability score for the industrial parks it serves.

\begin{table}[H]
    \caption{Scores for Top Candidate in Each Type of Highway Facility } \label{tab:highway_scores}
    \begin{center}
    \begin{tabular}{>{\raggedright\arraybackslash}m{0.15\linewidth}>{\centering\arraybackslash}m{0.075\linewidth}>{\centering\arraybackslash}m{0.07\linewidth}>{\centering\arraybackslash}m{0.07\linewidth}>{\centering\arraybackslash}m{0.07\linewidth}>{\centering\arraybackslash}m{0.07\linewidth}>{\raggedright\arraybackslash}m{0.07\linewidth}>{\raggedright\arraybackslash}m{0.08\linewidth}>{\raggedright\arraybackslash}m{0.08\linewidth}>{\raggedright\arraybackslash}m{0.08\linewidth}}
        \toprule[1.5pt]
        \multirow{2}{=}{\textbf{Candidate}} & \multicolumn{5}{c}{\textbf{Industrial Park} ($i$)} & \multicolumn{3}{c}{\textbf{Score}} \\ 
        \cmidrule(lr){2-6} \cmidrule(lr){7-9}
        & \multicolumn{1}{c}{\textbf{A}} & \multicolumn{1}{c}{\textbf{B}} & \multicolumn{1}{c}{\textbf{C}} & \multicolumn{1}{c}{\textbf{D}} & \multicolumn{1}{c}{\textbf{E}} &  $\Sigma \mathbf{x}_{i}^{\text{Score}}$ & $\mathbf{x}_{v}^\text{numBus}$  & \textbf{${Score}_v$} \\ 
        \midrule[1.5pt]
        E.Gunpo T\tnote{1} & \textcolor{blue}{Y} & \textcolor{blue}{Y} & \textcolor{red}{N} & \textcolor{red}{N} & \textcolor{red}{N} & 0.7185 & 213 & \textbf{153.05} \\ \hline
        Ansan R\tnote{2} & \textcolor{blue}{Y} & \textcolor{blue}{Y} & \textcolor{red}{N} & \textcolor{red}{N} & \textcolor{red}{N} & 0.7185 & 125 & \textbf{89.82} \\ \hline
        Guseong E\tnote{3} & \textcolor{blue}{Y} & \textcolor{red}{N} & \textcolor{red}{N} & \textcolor{blue}{Y} & \textcolor{red}{N} & 0.2978 & 478 & \textbf{142.36} \\ 
        \bottomrule[1.5pt]
    \end{tabular}
    \end{center}
\end{table}

\subsection{5.2 Characteristics of Top 10 Candidates}

\begin{figure}[H]
  \centering
   \includegraphics[width=0.75\textwidth]{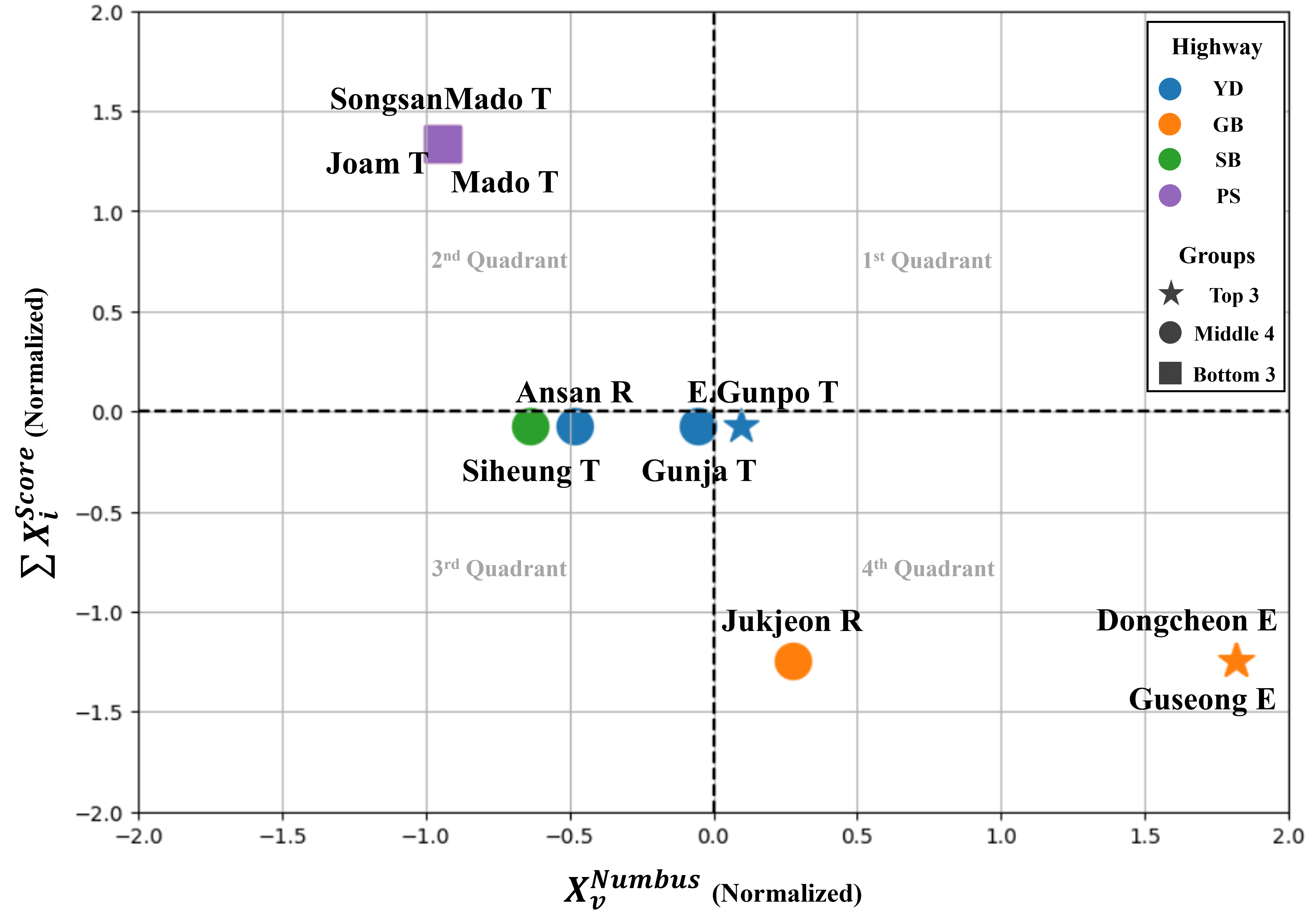}
  \caption{Quadrant Plot Based on the Number of Buses and Location Desirability}\label{figure:7}
\end{figure}

The top 10 candidates by total score (Table \ref{tab:Results_top10}) were analyzed using a quadrant plot (Figure \ref{figure:7}). This analysis considered the number of buses ($\mathbf{x}_{v}^\text{numBus}$) and the aggregated location desirability of each candidate's accessible destinations ($\Sigma \mathbf{x}_{i}^{\text{Score}}$).

\begin{table}[H]
    \centering
    \caption{Scores for Top 10 Candidates}
    \begin{center}
    \label{tab:Results_top10}
    \begin{tabular}{>{\raggedright\arraybackslash}m{0.14\linewidth}>{\raggedright\arraybackslash}m{0.08\linewidth}>{\raggedright\arraybackslash}m{0.09\linewidth}>{\raggedright\arraybackslash}m{0.075\linewidth}>{\raggedright\arraybackslash}m{0.14\linewidth}>{\raggedright\arraybackslash}m{0.08\linewidth}>{\raggedright\arraybackslash}m{0.09\linewidth}>{\raggedright\arraybackslash}m{0.075\linewidth}} 
    \toprule[1.5pt]
        \multicolumn{1}{c}{\textbf{$v$}} & \multicolumn{1}{c}{\textbf{$\mathbf{x}_{v}^\text{numBus}$}} & \multicolumn{1}{c}{\textbf{$\sum \mathbf{x}_{i}^\text{Score}$}} & \multicolumn{1}{c}{\textbf{${Score}_v$}} & \multicolumn{1}{c}{\textbf{$v$}} & \multicolumn{1}{c}{\textbf{$\mathbf{x}_{v}^\text{numBus}$}} & \multicolumn{1}{c}{\textbf{$\sum \mathbf{x}_{i}^\text{Score}$}} & \multicolumn{1}{c}{\textbf{${Score}_v$}} \\
        \midrule[1.5pt]
        E.Gunpo T\tnote{1} & 213 & 0.71853 & 153.04 & Siheung T & 101 & 0.71853 & 72.57 \\ \hline
        Guseong E\tnote{2} & 478 & 0.297831 & 142.36 & Jukjeon (S) R & 241 & 0.297831 & 71.77 \\ \hline
        Dongcheon E & 477 & 0.297831 & 142.06 & Mado T & 54 & 1.223338  & 66.06 \\ \hline
        Gunja T & 190 & 0.71853 & 136.52 & Songsan Mado T & 54 & 1.223338 & 66.06 \\ \hline
        Ansan R & 125 & 0.71853 & 89.81 & Joam T & 54 & 1.223338 & 66.06 \\
        \bottomrule[1.5pt]
    \end{tabular}
    \begin{tablenotes}
        \item[ ] \textit{T : Toll Gate, E : EX-HUB, R : Rest Area, ${Score}_v$ values are truncated to three decimal places}
    \end{tablenotes}
    \end{center}
\end{table} 

In Figure \ref{figure:7}, the term "Highway" in the legend refers to the abbreviations of the highway names to which each candidate is associated. Common characteristics among these candidates were identified through this approach. In the second quadrant, candidates are predominantly situated within designated industrial and production management areas (agriculture, forestry, and fisheries) based on cadastral maps. These candidates are associated with relatively low ${Score}_v$, all of which are among the bottom three in terms of ${\mathbf{x}_{v}^\text{numBus}}$. They are strategically positioned along the Pyeongtaek-Siheung(PS) Highway, ranked 27th in traffic volume among South Korean highways \cite{HighwayTraffic}, indicating a lower relative traffic volume. 
In the third quadrant, candidates are located within a mix of legal industrial zones, residential areas, and natural green spaces based on cadastral maps. The ${Score}_v$ for these candidates are moderate, indicating that they are neither in the highest nor the lowest categories, but rather reflect an intermediate level of ${\mathbf{x}_{v}^\text{numBus}}$.
In the fourth quadrant, this includes candidates located along the GyeongBu(GB) Highway, which has the heaviest traffic volume in South Korea, and the YeongDong(YD) Highway, ranked third in traffic volume \cite{HighwayTraffic}. The candidates are mixed within industrial and residential areas based on cadastral maps. These candidates have relatively high ${Score}_v$.

Consequently, candidates near highways with heavy traffic and residential areas are likely to experience increased bus traffic volume. This increase in bus traffic volume results in higher transfer efficiency, thereby enhancing the candidate's ${Score}_v$ within our proposed methodology.

\subsection{5.3 Characteristics of Top 2 Candidates based on the number of Bus-routes}
\begin{figure}[H]
  \centering
   \includegraphics[width=1\textwidth]{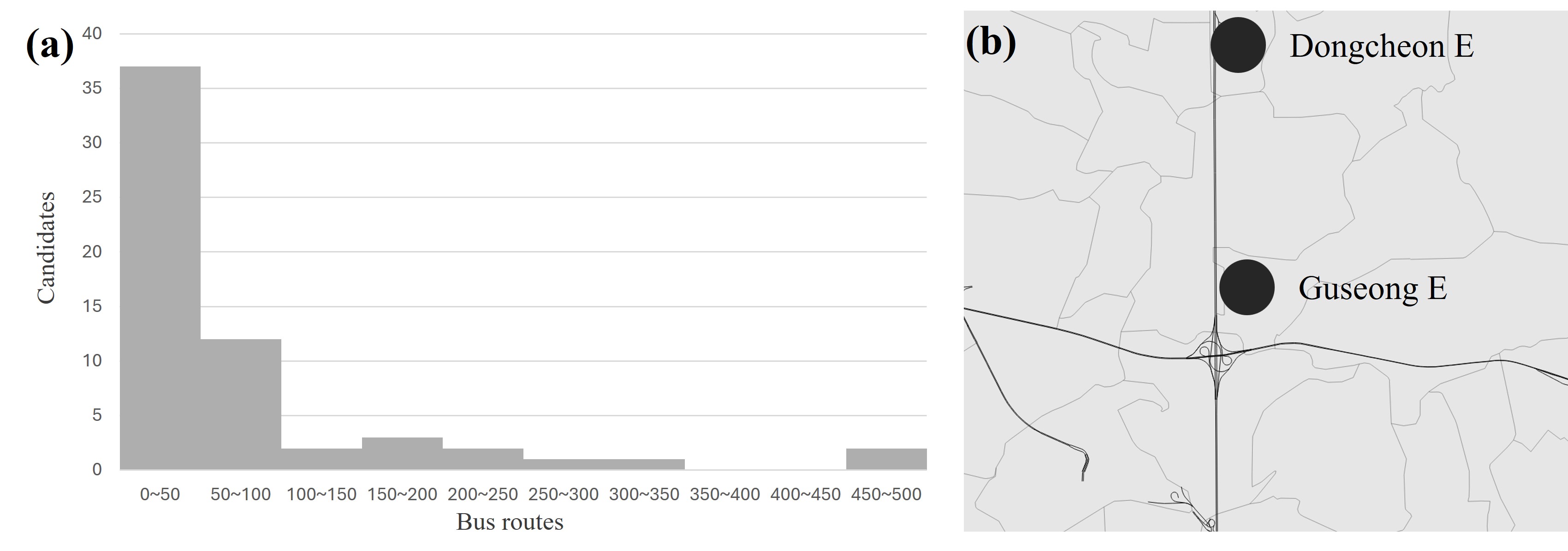}
  \caption{(a) Histogram of Bus Routes and (b) Locations of Dongcheon E and Guseong E }\label{figure:8}
\end{figure}

The number of bus routes passing through Vertiport candidates exceeds 400 at Guseong E (478 routes) and Dongcheon E (477 routes) as shown in Table \ref{tab:Results3}. In Figure \ref{figure:8} (a), these values fall into the top bin for all candidates in the histogram of bus routes. Considering their geographic location relative to highways as shown in Figure \ref{figure:8} (b), it is confirmed that these candidates are adjacent to the junctions of the Gyeongbu Highway (north-south axis) and Yeongdong Highway (west-east axis). Specifically, Guseong E is located 1 km and Dongcheon E is located 5 km from these junctions, suggesting these locations can absorb bus traffic volume from both highways.

\begin{table}[H]
    \centering
    \caption{Scores for Top 2 Candidates Based on the Number of Bus Routes}
    \begin{center}
    \label{tab:Results3}
    \begin{tabular}{>{\raggedright\arraybackslash}m{0.14\linewidth}>{\raggedright\arraybackslash}m{0.08\linewidth}>{\raggedright\arraybackslash}m{0.09\linewidth}>{\raggedright\arraybackslash}m{0.075\linewidth}>{\raggedright\arraybackslash}m{0.14\linewidth}>{\raggedright\arraybackslash}m{0.08\linewidth}>{\raggedright\arraybackslash}m{0.09\linewidth}>{\raggedright\arraybackslash}m{0.075\linewidth}} 
    \toprule[1.5pt]
        \multicolumn{1}{c}{\textbf{$v$}} & \multicolumn{1}{c}{\textbf{$\mathbf{x}_{v}^\text{numBus}$}} & \multicolumn{1}{c}{\textbf{$\sum \mathbf{x}_{i}^\text{Score}$}} & \multicolumn{1}{c}{\textbf{${Score}_v$}} & \multicolumn{1}{c}{\textbf{$v$}} & \multicolumn{1}{c}{\textbf{$\mathbf{x}_{v}^\text{numBus}$}} & \multicolumn{1}{c}{\textbf{$\sum \mathbf{x}_{i}^\text{Score}$}} & \multicolumn{1}{c}{\textbf{${Score}_v$}} \\
        \midrule[1.5pt]
        Guseong E\tnote{2} & 478 & 0.297831 & 142.36 & Dongcheon E & 477 & 0.297831 & 142.06  \\
        \bottomrule[1.5pt]
    \end{tabular}
    \begin{tablenotes}
        \item[ ] \textit{T : Toll Gate, E : EX-HUB, R : Rest Area, ${Score}_v$ values are truncated to three decimal places}
    \end{tablenotes}
    \end{center}
\end{table} 

\section{6. Conclusion and Future Study}
In this study, we aimed to propose a site selection methodology for integrating UAM with highway infrastructures, considering air operation availability, location desirability, and transfer effectiveness. This methodology strengthens the connection with the existing transportation network and improves vulnerable transportation areas. Also, based on our proposed methodology, we conducted an analysis focused on South Korea’s highway infrastructure and industrial parks to identify the optimal Vertiport locations using air operational filters and score function. As a result, our identified Vertiport locations reflect these characteristics by providing integrated transport options, considering weak public transportation networks at the destination and meeting the operational requirements of UAM. These findings suggest that integrating UAM with highway infrastructures can significantly enhance transportation efficiency and traveler satisfaction. This approach may also serve as a benchmark for other regions looking to implement similar transportation solutions, such as the Harbor Gateway Transit Center in the United States or Japan's Highway Oasis, a multifunctional transit rest area.

However, there were several limitations in this study. One limitation was the exclusion of real-time climate data, such as noise and fogging areas, which are critical for UAM operations. Another limitation was the focus on the number of bus routes passing through highway facilities without considering other important factors, such as the frequency of bus services and the characteristics of these routes.
Future research should focus on measuring the transfer effect by considering the frequency and volume of all transportation modes converging on nodes from the perspectives of (1) environmental data, (2) Mobility-as-a-Service (MaaS) and (3) willingness to transfer to UAM services.

\section{Acknowledgements}
This work was supported by National Research Foundation of Korea (NRF) grant funded by the Korea Government(MSIT) (No. 2022M3J6A1063021, No. 2022R1C1C2008155 and No. 2021R1A4A1033486). This work was also supported by Institute for Information \& communications Technology Planning \& Evaluation(IITP) grant funded by the Korea government(MSIT) (RS-2019-II190075, Artificial Intelligence Graduate School Program(KAIST)).

\newpage

\bibliographystyle{trb}
\bibliography{trb_template}
\end{document}